\documentclass[10pt,twocolumn,twoside]{IEEEtran}
\ifCLASSINFOpdf
\else
\fi
\usepackage{amsmath,amssymb,amsfonts}
\usepackage{cite}

\usepackage{algorithmic}
\usepackage{stfloats}
\usepackage{graphicx}
\usepackage{subfigure}
\usepackage{textcomp}
\usepackage{xcolor}
\usepackage{pifont}
\usepackage{amsthm}

\usepackage{mathrsfs}
\usepackage{cases}
\usepackage{comment}
\usepackage{empheq} 
\usepackage{caption}
\usepackage{bm}
\allowdisplaybreaks
\usepackage{xcolor}
\usepackage{url}  
\usepackage{graphicx}  
\usepackage[ruled,vlined,linesnumbered]{algorithm2e}
\usepackage{setspace}
\usepackage{floatpag} 
\usepackage{float}
\floatpagestyle{empty}

\renewcommand{\baselinestretch}{0.99}

\usepackage{listings}
\begin{document}

\title{Uplink Achievable Rate Maximization for Reconfigurable Intelligent Surface Aided Millimeter Wave Systems with Resolution-Adaptive ADCs }
%

%

\author{ 
Yue Xiu,~Jun Zhao,~\IEEEmembership{Member,~IEEE},
~Ertugrul Basar,~\IEEEmembership{Senior Member,~IEEE},~Marco Di Renzo,~\IEEEmembership{Fellow,~IEEE},~Wei Sun,~\IEEEmembership{Student Member,~IEEE}, ~Guan Gui,~\IEEEmembership{Senior Member,~IEEE},~Ning Wei,~\IEEEmembership{Member,~IEEE}\\
\thanks{Yue Xiu and Ning Wei are with University of Electronic Science and Technology of China, Chengdu, China (E-mail: xiuyue@std.uestc.edu.cn).
Jun Zhao is with School of Computer Science and Engineering, Nanyang Technological University, Singapore (E-mail: junzhao@ntu.edu.sg).
Ertugrul Basar is with Department of Electrical and Electronics Engineering
Koç University, Istanbul, Turkey (E-mail: ebasar@ku.edu.tr).
Marco Di Renzo is with Universit\'e Paris-Saclay, CNRS, CentraleSup\'elec, Laboratoire des Signaux et Syst\`emes, 91192 Gif-sur-Yvette, France. (E-mail: marco.direnzo@centralesupelec.fr). 
Wei Sun is with School of Computer Science and Engineering, Northeastern University, Shenyang 110819, China (E-mail: weisun@stumail.neu.edu.cn).
Guan Gui is with College of Telecommunications and Information Engineering, Nanjing University of Posts and Telecommunications, Nanjing, China (E-mail: guiguan@njupt.edu.cn).
}}

\maketitle

\begin{abstract}
In this letter, we investigate the uplink of a reconfigurable intelligent surface (RIS)-aided millimeter-wave (mmWave) multi-user system. In the considered system, however, problems with hardware cost and power consumption arise when massive antenna arrays coupled with power-demanding analog-to-digital converters (ADCs) are employed. To account for practical hardware complexity, we consider that the access point (AP) is equipped with resolution-adaptive analog-to-digital converters (RADCs). We maximize the achievable rate under hardware constraints by jointly optimizing the ADC quantization bits, the RIS phase shifts, and the beam selection matrix. Due to the non-convexity of the feasible set and objective function, the formulated problem is non-convex and difficult to solve. To efficiently tackle this problem, a block coordinated descent (BCD)-based algorithm is proposed. Simulations demonstrate that an RIS can mitigate the hardware loss due to use of RADCs, and that the proposed BCD-based algorithm outperforms state-of-the-art algorithms.

\end{abstract}


%
\IEEEpeerreviewmaketitle

\section{Introduction}
Millimeter-wave (mmWave) communication systems play an important role in fifth generation (5G) wireless networks. MmWave communication systems can offer a higher transmission capacity compared with their microwave counterpart. However, they are impaired by blockages, which affect their reliability, especially in urban environments characterized by the presence of large and densely deployed buildings \cite{b1},\cite{b2}.  

To enhance the reliability of mmWave communication systems, several solutions can be employed \cite{b2}. Recently, the emerging technology of reconfigurable intelligent surfaces (RISs) has been proposed for enhancing the system performance, especially at high frequency bands \cite{b1}. In particular, by appropriately co-phasing the incident signals, RISs provide high beamforming gains that increase the coverage of mmWave communication systems, without the need of using power amplifiers and multiple radio frequency (RF) chains that are, on the other hand, needed if relays are employed \cite{b3,b3_1,b6,b17,b18}.

Recently, RIS-aided communication systems have been investigated in several works. 
In \cite{b9_2}, the authors proposed a projected gradient method (PGM) for maximizing the achievable rate of RIS-aided multiple-input multiple-output (MIMO) systems. In \cite{b9_1}, the authors analyzed the capacity of RIS-aided indoor mmWave systems and studied effectiveness of RISs to alleviate the impact of blockages. 
In \cite{b10}, the authors analyzed the joint passive and hybrid beamforming optimization of mmWave systems, and a low-complexity iterative algorithm was proposed. 

In mmWave communication systems, the prohibitive cost and power consumption of the hardware components at the access points (APs) make the realization of fully-digital solutions very difficult. For these reasons, hybrid analog-digital processing schemes are usually employed to reduce the number of RF chains \cite{b10}. In addition, low-resolution analog-to-digital converters (ADCs) are often employed in order to further reduce the hardware cost and power consumption. In particular, a viable solution to find a good trade-off between hardware complexity, power consumption, cost, and performance is to use resolution adaptive ADCs (RADCs) \cite{b11}.

To the best of our knowledge, no research work has yet investigated the design and optimization of RIS-aided mmWave communication systems with RADCs. Therefore, it is meaningful to study the impact of using RADCs for application to RIS-aided uplink mmWave communication systems. Specifically, we  consider the problem of maximizing the achievable rate by jointly optimizing the beam selection matrix at the AP, the RIS phase shifts, and the ADC quantization bits at the AP. The resulting optimization problem is non-convex, and, thus, it is difficult to solve. To circumvent this issue,   we   propose   a   block   coordinated   descent   (BCD)-based  algorithm. Simulations reveal that by dynamically adjusting the resolution of the RADCs, and the phase shifts of the RIS, the hardware loss can be mitigated and the throughput of RIS-aided mmWave systems can be improved.

\section{System Model and Problem Formulation}
\begin{figure}[!t]
\centering
\includegraphics[height=1.2in,width=1.9in]{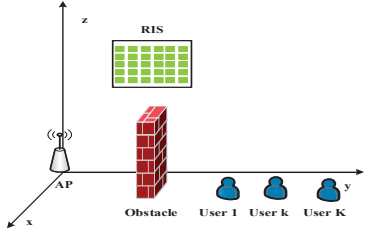}
\caption{RIS-aided uplink mmWave communication system.}
\label{fig:1}
\end{figure}
We consider a multi-user uplink mmWave communication system, where $K$ single-antenna users are served by an AP with $M$ RF chains and $N$ antennas. To assist the communication between the AP and the users, an RIS equipped with $N_{r}$ reflecting elements is assumed to be available, as illustrated in Fig. 1. We adopt a geometric model for the mmWave channels \cite{b10}. In particular, $\boldsymbol{h}_{k}\in\mathbb{C}^{N_{r}\times 1}$ denotes the channel from the $k$th user to the RIS, and $\boldsymbol{H}=[\boldsymbol{h}_{1},\ldots,\boldsymbol{h}_{K}]\in\mathbb{C}^{N_{r}\times K}$ is the channel matrix that accounts for the $K$ users, $\boldsymbol{s}=[s_{1},\ldots,s_{K}]^{T}\in\mathbb{C}^{K\times 1}$  with $s_{k}\sim\mathcal{CN}(0,1)$ denotes the $k$th user's data, $\boldsymbol{\Theta}=\mathrm{diag}(\boldsymbol{\theta})\in\mathbb{C}^{N_{r}\times N_{r}}$ denotes the RIS reflection matrix, where $\boldsymbol{\theta}=[\beta e^{j\phi_{1}},\ldots, \beta e^{j\phi_{
N_{r}}}]\in\mathbb{C}^{1\times N_{r}}$, $\phi_{i}\in [0,2\pi],~\forall~i=1,\ldots,N_{r}$, and $\beta=1$ denote the phase shift and the amplitude reflection coefficient of the $i$th reflecting element of the RIS, respectively.
Due to the presence of large objects, e.g., buildings, the direct links from the users to the AP are assumed to be weak, and, therefore, are ignored.
The signal received by the AP through the RIS can be expressed as 
\begin{align}
\boldsymbol{y}=\boldsymbol{G}\boldsymbol{\Theta}\boldsymbol{H}\boldsymbol{s}+\boldsymbol{n},\label{1}
\end{align}
where $\boldsymbol{n}\sim\mathcal{CN}(0,\sigma^{2}\boldsymbol{I}_{N})$ is the additive white Gaussian noise vector and $\boldsymbol{I}_{N}$ denotes an $N \times N$ identity matrix, and $\boldsymbol{G}\in\mathbb{C}^{N\times N_{r}}$ is the channel matrix between the RIS and the AP. The received signal $\boldsymbol{y}\in\mathbb{C}^{N\times 1}$ is processed by the AP by applying a hybrid combiner with RADCs. The hybrid combiner is denoted by $\boldsymbol{F}=\boldsymbol{D}\boldsymbol{W}\in\mathbb{C}^{N\times M}$, where $\boldsymbol{D}\in\mathbb{C}^{N\times S}$ is a codebook 
matrix, and
$\boldsymbol{W}\in\mathbb{C}^{S\times M}$ denotes a selection matrix with binary entries $w_{s,m}\in\{0,1\}$. After applying the hybrid combiner, the signal is denoted as
\begin{align}
\bar{\boldsymbol{y}}=\boldsymbol{F}^{H}\boldsymbol{G}\boldsymbol{\Theta}\boldsymbol{H}\boldsymbol{s}+\boldsymbol{F}^{H}\boldsymbol{n}.\label{2}
\end{align}
The $M$ pairs RADCs are assumed to be connected to the RF processor in order to make the control of the quantization bits more flexible and precise, thereby reducing the allocation of the quantization error. The quantization noise of the RADCs for the real and imaginary parts of $\boldsymbol{y}$ and $\bar{\boldsymbol{y}}$ is taken into account by employing a linear
additive quantization noise model (AQNM) \cite{b16}.
Thus, the quantized signal can be formulated as
\begin{align}
\tilde{\boldsymbol{y}}=\mathcal{F}(\bar{\boldsymbol{y}})=\boldsymbol{F}_{\alpha}\bar{\boldsymbol{y}}+\boldsymbol{n}_{q},\label{3}
\end{align}
where $\mathcal{F}(\cdot)$ denotes the quantization operator and $\boldsymbol{F}_{\alpha}=\alpha\boldsymbol{I}_{M}\in\mathbb{C}^{M\times M}$, where $\boldsymbol{I}_{M}$ is an $M \times M$ identity matrix and $\alpha=\frac{\pi\sqrt{3}}{2}4^{-b}$ denotes the normalized quantization error for $b$ quantization bits~\cite{b16}. Also, $\boldsymbol{n}_{q}$ denotes the quantized noise, whose mean and covariance matrix are $0$ and $\boldsymbol{A}_{a}=\boldsymbol{F}_{\alpha}\boldsymbol{F}_{b}\mathrm{diag}(\boldsymbol{F}^{H}\boldsymbol{G}\boldsymbol{\Theta}\boldsymbol{H}\boldsymbol{H}^{H}\boldsymbol{G}^{H}\boldsymbol{\Theta}^{H}\boldsymbol{F}+\sigma^{2}\boldsymbol{F}^{H}\boldsymbol{F})$, respectively, with $\boldsymbol{F}_{b}=b\boldsymbol{I}_{M}\in\mathbb{C}^{M\times M}$. According to this model, the detected signal of the $k$th user is given by
\begin{align}
\hat{s}_{k}=\boldsymbol{u}_{k}^{H}\boldsymbol{F}_{\alpha}\boldsymbol{F}^{H}\boldsymbol{G}\boldsymbol{\Theta}\boldsymbol{H}\boldsymbol{x}+\boldsymbol{u}_{k}^{H}\boldsymbol{F}_{\alpha}\boldsymbol{F}^{H}\boldsymbol{n}+\boldsymbol{u}_{k}^{H}\boldsymbol{n}_{q},\label{4}
\end{align}
where $\boldsymbol{u}_k$ denotes the decoding vector of the $k$th user.
For convenience, we define $\boldsymbol{w}=\mathrm{vec}(\boldsymbol{W})\in\mathbb{C}^{SM\times 1}$ and $\boldsymbol{u}=[\boldsymbol{u}_{1}^{T},\ldots,\boldsymbol{u}_{K}^{T}]^{T}\in\mathbb{C}^{MK\times 1}$. Based on this model, the achievable rate of the $k$th user is given in (5), shown in the next page.
\newcounter{mytempeqncnt}
\begin{figure*}[!t]
\normalsize
\setcounter{mytempeqncnt}{\value{equation}}
\begin{eqnarray}
{}{R_{k}=\log\left(1+\frac{|\boldsymbol{u}_{k}^{H}\boldsymbol{F}_{\alpha}\boldsymbol{F}^{H}\boldsymbol{G}\boldsymbol{\Theta}\boldsymbol{h}_{k}|^{2}}{\sum\nolimits_{l\neq k}|\boldsymbol{u}_{k}^{H}\boldsymbol{F}_{\alpha}\boldsymbol{F}^{H}\boldsymbol{G}\boldsymbol{\Theta}\boldsymbol{h}_{l}|^{2}+\sigma^{2}\|\boldsymbol{u}_{k}^{H}\boldsymbol{F}_{\alpha}\boldsymbol{F}^{H}\|^{2}+\boldsymbol{u}_{k}^{H}\boldsymbol{A}_{a}\boldsymbol{u}_{k}}\right)}.
\end{eqnarray}
\setcounter{equation}{31}
\begin{eqnarray}
{}{\boldsymbol{B}_{k}=\sum\nolimits_{l=1}^{K}\boldsymbol{F}_{\alpha}\boldsymbol{F}^{H}\boldsymbol{G}\boldsymbol{\Theta}\boldsymbol{h}_{l}\boldsymbol{h}_{l}^{H}\boldsymbol{\Theta}^{H}\boldsymbol{G}^{H}\boldsymbol{F}\boldsymbol{F}_{\alpha}^{H}+\sigma^{2}\boldsymbol{F}_{\alpha}\boldsymbol{F}^{H}\boldsymbol{F}\boldsymbol{F}_{\alpha}^{H}+\boldsymbol{A}_{a}}.\label{32}
\end{eqnarray}
\begin{eqnarray}
{}{\boldsymbol{D}_{k}=\sum\nolimits_{l\neq k}\boldsymbol{F}_{\alpha}\boldsymbol{F}^{H}\boldsymbol{G}\boldsymbol{\Theta}\boldsymbol{h}_{l}\boldsymbol{h}_{l}^{H}\boldsymbol{\Theta}^{H}\boldsymbol{G}^{H}\boldsymbol{F}\boldsymbol{F}_{\alpha}^{H}+\sigma^{2}\boldsymbol{F}_{\alpha}\boldsymbol{F}^{H}\boldsymbol{F}\boldsymbol{F}_{\alpha}^{H}+\boldsymbol{A}_{a}}.\label{33_1}
\end{eqnarray}
\setcounter{equation}{\value{mytempeqncnt}}
\hrulefill
\vspace*{4pt}
\end{figure*}

We are interested in optimizing the phase shifts of the RIS $\boldsymbol{\Theta}$, the allocation of the quantization bits $b$ of the RADCs, the selection matrix $\boldsymbol{w}$, and the decoding vector $\boldsymbol{u}$ to maximize the uplink achievable rate. This problem can be formulated as
\setcounter{equation}{5}
\begin{subequations}
\begin{align}
{}{\max\limits_{\boldsymbol{\Theta},b,\boldsymbol{u},\boldsymbol{w}}}~&{}{\sum\nolimits_{k=1}^{K}R_{k}}\label{6a}\\
\mbox{s.t.}~
&\sum\nolimits_{s=1}^{S}w_{s,m}=1,~\sum\nolimits_{m=1}^{M}w_{s,m}\leq 1,~w_{s,m}\in\{0,1\},&\nonumber\\ 
&\forall~s,m,&\label{6b}\\
&{}{|\theta_{i}|=1,}&\label{6c}\\
&b^{\min}\leq b \leq b^{\max},~~b~\text{is~an~integer}.&\label{6d}
\end{align}\label{6}
\end{subequations}\vspace{-20pt}

\section{Proposed BCD Algorithm}
\subsection{Optimization of $b$ and $\boldsymbol{w}$}
For given $\boldsymbol{u}$ and $\boldsymbol{\Theta}$, the problem in (\ref{6}) is rewritten as
\begin{subequations}
\begin{align}
{}{\max\limits_{b,\boldsymbol{w}}}~&{}{\sum\nolimits_{k=1}^{K}R_{k}}\label{7a}\\
\mbox{s.t.}~
&\text{(\ref{6b})}, \text{(\ref{6d})}.&\label{7b}
\end{align}\label{7}
\end{subequations}
The problem in (\ref{7}) is non-convex due to the nonconvexity of (\ref{6b}) and (\ref{6d}). Therefore,
to handle the non-convex discrete constraint (\ref{6d}), we relax constraint (\ref{6d}) into a continuous constraint, i.e.,
\begin{align}
{}{b^{\min}\leq b\leq b^{\max}}.\label{8}
\end{align}
According to \cite{b11}, $b$ is rounded as
\begin{align}
{}{b(\delta)=\begin{cases}
\lfloor b^{*}\rfloor, \text{if}~~b^{*}-\lfloor b^{*}\rfloor\leq\delta \\
\lceil b^{*}\rceil,  \text{otherwise},
\end{cases}}\label{9}
\end{align}
where $0\leq \delta\leq 1$ is chosen based on \cite{b11}. In addition, in order to address the difficulties caused by constraint (\ref{6b}), an appropriate transformation is required. In particular, (\ref{6b}) can be transformed into the following equivalent form
\begin{align}
{}{\boldsymbol{w}_{s}^{T}\boldsymbol{e}_{m}=\hat{w}_{s,m}, \sum\nolimits_{s=1}^{S}\boldsymbol{w}_{s}^{T}\boldsymbol{e}_{m}=1,}\label{10}
\end{align}
\begin{align}
{}{\boldsymbol{w}_{s}^{T}\boldsymbol{e}_{m}(1-\hat{w}_{s,m})=0},\label{11}
\end{align}
\begin{align}
{}{\boldsymbol{w}_{s}\succeq \boldsymbol{0},~\boldsymbol{w}_{s}^{T}\boldsymbol{1}_{M}\leq 1,~0\leq \hat{w}_{s,m}\leq 1},\label{12}
\end{align}
where $\boldsymbol{e}_{m}=\boldsymbol{I}(:,m)$ and $\boldsymbol{1}_{M}=[1,\ldots,1]^{T}\in\mathbb{R}^{M\times 1}$. To deal with the bilinear variables, $\boldsymbol{w}_{s}^{T}\boldsymbol{e}_{m}(1-\hat{w}_{s,m})=0$ can be transformed into the following constraints by the Schur complement \cite{b15}
\begin{align}
{}{\left[
 \begin{matrix}
   \boldsymbol{w}_{s}^{T}\boldsymbol{e}_{m} & r_{s,m} \\
   r_{s,m} & 1-\hat{w}_{s,m}
  \end{matrix}
  \right]\succeq\boldsymbol{0}},\label{13}
\end{align}
and
\begin{align}
{}{\boldsymbol{w}_{s}^{T}\boldsymbol{e}_{m}(1-\hat{w}_{s,m})\leq r_{s,m}^{2}},\label{14}
\end{align}
where $r_{s,m}$ is an auxiliary variable. In  order  to deal  with  the  bilinear  function  on the left-hand side of (\ref{14}),  the  sequential convex approximation (SCA)  method  based  on the arithmetic  geometric  mean  (AGM) inequality is adopted. Accordingly, (\ref{14}) can be rewritten as
\begin{align}
&\boldsymbol{w}_{s}^{T}\boldsymbol{e}_{m}(1-\hat{w}_{s,m})\leq \frac{1}{2}\left((\boldsymbol{w}_{s}^{T}\boldsymbol{e}_{m}\eta_{s,m})^{2}+(\frac{1-\hat{w}_{s,m}}{\eta_{s,m}})^{2}\right)\nonumber\\
&\leq r_{s,m}^{2}, \label{15}   
\end{align}
where $\eta_{s,m}$ is a feasible point. To tighten the upper bound, $\eta_{s,m}$ is iteratively updated. In particular, at the $n$-th iteration $\eta_{s,m}$ is expressed as
\begin{align}
{}{\eta_{s,m}^{(n)}=\sqrt{(1-\hat{w}_{s,m}^{(n-1)})/((\boldsymbol{w}_{s}^{T})^{(n-1)}\boldsymbol{e}_{m})}.}\label{16}
\end{align}
However, the constraint in (\ref{15}) is still non-convex. Then, we use the SCA method to transform (\ref{15}) into the following convex constraint
\begin{align}
{}{\frac{1}{2}\left((\boldsymbol{w}_{s}^{T}\boldsymbol{e}_{m}\eta_{s,m})^{2}+(\frac{1-\hat{w}_{s,m}}{\eta_{s,m}})^{2}\right)-\bar{r}_{s,m}(r_{s,m}-\bar{r}_{s,m})\leq 0.} \label{17} 
\end{align}

Moreover, by introducing the auxiliary variables $\omega_{k}$ and $\zeta$ to deal with the non-convex objective function in (\ref{7a}), $R_{k}$ is rewritten as
\begin{align}
{}{R_{k}=\log\left(1+\zeta|\boldsymbol{w}\boldsymbol{A}_{k}|^{2}/\omega_{k}\right)},\label{18}
\end{align}
where 
\begin{align}
&\omega_{k}=\sum\nolimits_{l\neq k}|\boldsymbol{u}_{k}^{H}\boldsymbol{F}_{\alpha}\boldsymbol{F}^{H}\boldsymbol{G}\boldsymbol{\Theta}\boldsymbol{h}_{l}|^{2}+\sigma^{2}\|\boldsymbol{u}_{k}^{H}\boldsymbol{F}_{\alpha}\boldsymbol{F}^{H}\|^{2}\nonumber\\
&+\boldsymbol{u}_{k}^{H}\boldsymbol{A}_{a}\boldsymbol{u}_{k},\label{19}\\
&{}{\bm{A}_{k}=((\boldsymbol{D}^{H}\boldsymbol{G}\boldsymbol{\Theta}\boldsymbol{h}_{k})^{T}\otimes(\boldsymbol{u}_{k}^{H}\boldsymbol{F}_{\alpha}))^{H}},\label{20}\\
&{}{\log_{4}((\pi\sqrt{3})/(2\zeta))-b=0}.\label{21}
\end{align}
To tackle the non-convex constraint in (\ref{21}), we use the SCA method to transform (\ref{21}) in the following convex constraints
\begin{align}
b+\frac{2\bar{\zeta}}{\pi\sqrt{3}}\ln_{4}\frac{\pi\sqrt{3}}{2\bar{\zeta}^{2}}(\zeta-\bar{\zeta})\leq 0, \log_{4}(\frac{\pi\sqrt{3}}{2\zeta})-b\geq 0.\label{22}
\end{align}
Due  to  the  coupling  between $\zeta$ and $\boldsymbol{w}$, however, the objective function in (\ref{18})  is still  intractable.  To obtain a tractable problem, (\ref{18}) is transformed to
\begin{align}
{}{R_{k}=\log(1+\rho_{k})},\label{23}
\end{align}
where 
\begin{align}
\rho_{k}\leq \zeta|\boldsymbol{w}\boldsymbol{A}_{k}|^{2}/\omega_{k}.\label{24}
\end{align}
By using a similar line of thought, we have
\begin{align}
{}{\left[
 \begin{matrix}
   \zeta & t_{k} \\
   t_{k} & |\boldsymbol{w}\boldsymbol{A}_{k}|^{2}
  \end{matrix}
  \right]\succeq\boldsymbol{0}},\label{25}
\end{align}

\begin{align}
t_{k}^{2}/\omega_{k}\geq \rho_{k}.\label{26}
\end{align}
Then,  we  use  the  SCA  method  based  on  the  first-order Taylor expansion to tackle (\ref{26}). Specifically, the left-hand side of (\ref{26}) is non-convex with respect to $t_{k}$ and $\omega_{k}$, and thus it can be tightly bounded from below with its first-order Taylor approximation. In particular, for any fixed points $(\bar{t}_{k}, \bar{\omega}_{k})$, we have
\begin{align}
{}{\frac{t_{k}^{2}}{\omega_{k}}\geq \frac{2\bar{t}_{k}}{\bar{\omega}_{k}}t_{k}-\frac{\bar{t}_{k}^{2}}{\bar{\omega}_{k}^{2}}\omega_{k}\geq \rho_{k}}.\label{27}
\end{align}
By applying the SCA method in \cite{b6}, we iteratively update $\bar{t}_{k}$ and $\bar{\omega}_{k}$ at the $n$th iteration as
\begin{align}
{}{\bar{\omega}_{k}^{(n)}=\omega_{k}^{(n-1)}}, 
\bar{t}_{k}^{(n)}=t_{k}^{(n-1)}.\label{28}
\end{align}
Therefore, the problem in (7) is transformed into the following convex problem
\begin{subequations}
\begin{align}
{}{\max\limits_{b,\boldsymbol{w}, r_{s,m}, \hat{w}_{s,m}, \rho_{k}, \zeta, t_{k}}}&~{}{\sum\nolimits_{k=1}^{K}\log_{2}(1+\rho_{k})}\label{29a}\\
\mbox{s.t.}&~\text{(\ref{10})},\text{(\ref{12})},\text{(\ref{13})}, \text{(\ref{17})}, \text{(\ref{22})}, \text{(\ref{25})}, \text{(\ref{27})}.&\label{29b}
\end{align}\label{29}%
\end{subequations}
The SCA-based algorithm is given in \textbf{Algorithm~1}.
\begin{algorithm}[htbp]
\caption{SCA-based algorithm for solving problem (\ref{7}).} 
\hspace*{0.02in}{\bf Initialization:} $\bar{r}_{s,m}$, $\bar{\omega}_{k}$, $\bar{t}_{k}$, $\forall~k$.\\
\hspace*{0.02in}{\bf Repeat}\\
Update $\{b^{(n)},\boldsymbol{w}^{(n)}, r_{s,m}^{(n)}, \hat{w}_{s,m}^{(n)}, \rho_{k}^{(n)}, \zeta^{(n)}, t_{k}^{(n)}\}$ with fixed $\bar{r}_{s,m}$, $\bar{\omega}_{k}$, $\bar{t}_{k}$ by solving (\ref{29}).\\
Update $\eta_{s,m}^{(n)}$, $\bar{\omega}_{k}^{(n+1)}$, $\bar{t}_{k}^{(n+1)}$ based on (\ref{16}) and (\ref{28}).\\
Update $n=n+1$.\\
\hspace*{0.02in}{\bf Until}
Convergence.\\
\hspace*{0.02in}{\bf Output:} 
$\boldsymbol{w}^{*}, b^{*}$.\\
\end{algorithm}

\subsection{Optimization of $\boldsymbol{u}$ and $\boldsymbol{\Theta}$}
For given $b$, $\boldsymbol{w}$, $\boldsymbol{\Theta}$, the problem in (\ref{6}) is rewritten as
\begin{align}
{}{\max\limits_{\boldsymbol{u}_{k}}}~&{}{\sum\nolimits_{k=1}^{K}R_{k}}.\label{30}
\end{align}
Since the achievable rate of the $k$th user is only related to the decoding vector of the $k$th user, maximizing $\sum\nolimits_{k=1}^{K}R_{k}$ is equivalent to maximizing $R_{k}$ of each user by optimizing the decoding vector $\boldsymbol{u}_{k}$ of each user.
Therefore, the problem in (\ref{30}) can be recast as
\begin{align}
{}{\max_{\boldsymbol{u}_{k}}}~&{}{\frac{\boldsymbol{u}_{k}^{H}\boldsymbol{B}_{k}\boldsymbol{u}_{k}}{\boldsymbol{u}_{k}^{H}\boldsymbol{D}_{k}\boldsymbol{u}_{k}},~~\forall~k},\label{31}
\end{align}%
where $\boldsymbol{B}_{k}$ and $\boldsymbol{D}_{k}$ are given in (32) and (33) at the top of this page. 
We use the majorize-minimization (MM) algorithm to solve (\ref{31}) \cite{b13}. Specifically, 
let $y=\boldsymbol{u}_{k}^{H}\boldsymbol{D}_{k}\boldsymbol{u}_{k}$,  $h(\boldsymbol{u}_{k})=\frac{\boldsymbol{u}_{k}^{H}\boldsymbol{B}_{k}\boldsymbol{u}_{k}}{y}$ is jointly convex in $\boldsymbol{u}_{k}$ and $y$ because $\boldsymbol{B}_{k}$ is positive definite. Thanks to the convexity, we have
\setcounter{equation}{33}
\begin{align}
&h(\boldsymbol{u}_{k})=\frac{\boldsymbol{u}_{k}^{H}\boldsymbol{B}_{k}\boldsymbol{u}_{k}}{\boldsymbol{u}_{k}^{H}\boldsymbol{D}_{k}\boldsymbol{u}_{k}}\geq 2\frac{\mathrm{Re}(\bar{\boldsymbol{u}}_{k}^{H}\boldsymbol{B}_{k}\boldsymbol{u}_{k})}{\bar{\boldsymbol{u}}_{k}^{H}\boldsymbol{D}_{k}\bar{\boldsymbol{u}}_{k}}-\frac{\bar{\boldsymbol{u}}_{k}^{H}\boldsymbol{B}_{k}\bar{\boldsymbol{u}}_{k}}{(\bar{\boldsymbol{u}}_{k}^{H}\boldsymbol{D}_{k}\bar{\boldsymbol{u}}_{k})^{2}}\nonumber\\
&\boldsymbol{u}_{k}^{H}\boldsymbol{D}_{k}\boldsymbol{u}_{k}+c.\label{34}
\end{align}
Based on Lemma 1 in\cite{b13}, a lower bound of  $\boldsymbol{u}_{k}^{H}\boldsymbol{D}_{k}\boldsymbol{u}_{k}$ is 
\begin{align}
{}{\boldsymbol{u}_{k}^{H}\boldsymbol{D}_{k}\boldsymbol{u}_{k}}&{}{\leq \boldsymbol{u}_{k}^{H}\lambda_{max}(\boldsymbol{D}_{k})\boldsymbol{u}_{k}+
2\mathrm{Re}(\boldsymbol{u}_{k}^{H}(\boldsymbol{D}_{k}-\lambda_{max}(\boldsymbol{D}_{k})\boldsymbol{I})}\nonumber\\
&{}{\bar{\boldsymbol{u}}_{k})},\label{35}
\end{align}
where $\lambda_{max}(\boldsymbol{D}_{k})$ is the maximum eigenvalue of matrix $\boldsymbol{D}_{k}$. Substituting (\ref{35}) into (\ref{34}), we have
\begin{align}
{}{\frac{\boldsymbol{u}_{k}^{H}\boldsymbol{B}_{k}\boldsymbol{u}_{k}}{\boldsymbol{u}_{k}^{H}\boldsymbol{D}_{k}\boldsymbol{u}_{k}}}&{}{\geq2\frac{\mathrm{Re}(\bar{\boldsymbol{u}}_{k}^{H}\boldsymbol{B}_{k}\boldsymbol{u}_{k})}{\bar{\boldsymbol{u}}_{k}^{H}\boldsymbol{D}_{k}\bar{\boldsymbol{u}}_{k}}-\frac{\bar{\boldsymbol{u}}_{k}^{H}\boldsymbol{B}_{k}\bar{\boldsymbol{u}}_{k}}{(\bar{\boldsymbol{u}}_{k}^{H}\boldsymbol{D}_{k}\bar{\boldsymbol{u}}_{k})^{2}}(\boldsymbol{u}_{k}^{H}\lambda_{max}(\boldsymbol{D}_{k})}\nonumber\\
&{}{\boldsymbol{u}_{k}+2\mathrm{Re}(\boldsymbol{u}_{k}^{H}(\boldsymbol{D}_{k}-\lambda_{max}(\boldsymbol{D}_{k})\boldsymbol{I})\bar{\boldsymbol{u}}_{k})\geq g(\boldsymbol{u}_{k}|\bar{\boldsymbol{u}}_{k})}\nonumber\\
&{}{
+[h(\bar{\boldsymbol{u}}_{k})-g(\bar{\boldsymbol{u}}_{k}|\bar{\boldsymbol{u}}_{k})]},\label{36}
\end{align}
where 
\begin{align}
{}{g(\boldsymbol{u}_{k}|\bar{\boldsymbol{u}}_{k})}&={}{2\frac{\mathrm{Re}(\bar{\boldsymbol{u}}_{k}^{H}\boldsymbol{B}_{k}\boldsymbol{u}_{k})}{\bar{\boldsymbol{u}}_{k}^{H}\boldsymbol{D}_{k}\bar{\boldsymbol{u}}_{k}}-\frac{\bar{\boldsymbol{u}}_{k}^{H}\boldsymbol{B}_{k}\bar{\boldsymbol{u}}_{k}}{(\bar{\boldsymbol{u}}_{k}^{H}\boldsymbol{D}_{k}\bar{\boldsymbol{u}}_{k})^{2}}(\boldsymbol{u}_{k}^{H}\lambda_{max}(\boldsymbol{D}_{k})}\nonumber\\
&{}{\boldsymbol{u}_{k}+2\mathrm{Re}(\boldsymbol{u}_{k}^{H}(\boldsymbol{D}_{k}-\lambda_{max}\boldsymbol{D}_{k})\bar{\boldsymbol{u}}_{k})}.\label{37}
\end{align}
\begin{align}
{}{h(\bar{\boldsymbol{u}}_{k})}=(\bar{\boldsymbol{u}}_{k}^{H}\boldsymbol{B}_{k}\bar{\boldsymbol{u}}_{k})/(\bar{\boldsymbol{u}}_{k}^{H}\boldsymbol{D}_{k}\bar{\boldsymbol{u}}_{k}).\label{37_1}
\end{align}
Since $h(\bar{\boldsymbol{u}}_{k})-g(\bar{\boldsymbol{u}}_{k}|\bar{\boldsymbol{u}}_{k})$ is a constant, the decoding vector optimization problem in each iteration of the MM algorithm is equivalent to solving the following problem
\begin{align}
{}{\max_{\boldsymbol{u}_{k}}}&~{}{g(\boldsymbol{u}_{k}|\bar{\boldsymbol{u}}_{k})=\mathrm{Re}[(\boldsymbol{v}^{(t)})^{H}\boldsymbol{u}_{k}-\beta^{(t)}\boldsymbol{u}_{k}^{H}\boldsymbol{u}_{k}]},\label{38}
\end{align}
where 
\begin{align}
&\boldsymbol{v}^{(t)}=\frac{\boldsymbol{B}_{k}\bar{\boldsymbol{u}}_{k}^{(t)}}{(\bar{\boldsymbol{u}}_{k}^{(t)})^{H}\boldsymbol{D}_{k}\bar{\boldsymbol{u}}_{k}^{(t)}}-\frac{(\bar{\boldsymbol{u}}_{k}^{(t)})^{H}\boldsymbol{B}_{k}\bar{\boldsymbol{u}}_{k}^{(t)}[\boldsymbol{D}_{k}
-\lambda_{max}(\boldsymbol{D}_{k})\boldsymbol{I}]}{[(\bar{\boldsymbol{u}}_{k}^{(t)})^{H}\boldsymbol{D}_{k}\bar{\boldsymbol{u}}_{k}^{(t)}]^{2}}\nonumber\\
&\times\bar{\boldsymbol{u}}_{k}^{(t)},\label{39}
\end{align}
\begin{align}
\beta^{(t)}=(\lambda_{max}(\boldsymbol{D}_{k})\bar{\boldsymbol{u}}_{k}^{H}\boldsymbol{B}_{k}\bar{\boldsymbol{u}}_{k})/(\bar{\boldsymbol{u}}_{k}^{H}\boldsymbol{D}_{k}\bar{\boldsymbol{u}}_{k})^{2}.\label{39_1}
\end{align}
By checking the first-order optimality
condition of (\ref{39}), we have
\begin{align}
{}{\boldsymbol{u}_{k}^{(t+1)}=\boldsymbol{v}^{(t)}/\beta^{(t)}.}\label{39_2}
\end{align}
The MM-based algorithm is given in \textbf{Algorithm~2}
\begin{algorithm}[htpb]
\caption{MM-based algorithm for (\ref{30}).} 
\hspace*{0.02in}{\bf Initialization:} $\bar{\boldsymbol{u}}_{k}^{(0)}$, $t=0$.\\
\hspace*{0.02in}{\bf Repeat:}\\
Update $\boldsymbol{u}_{k}^{(t+1)}$ based on (\ref{39_2}).
\\
Update $t=t+1$\\
\hspace*{0.02in}{\bf Until:}
Convergence.\\
\hspace*{0.02in}{\bf Output:}
the solution $\boldsymbol{u}_{k}^{*}$
\end{algorithm}.

We are then left with the optimization of $\boldsymbol{\Theta}$. For given $b$, $\boldsymbol{u}_{k}$ and $\boldsymbol{W}$, (\ref{6}) is rewritten as
\begin{subequations}
\begin{align}
{}{\max\limits_{\boldsymbol{\Theta}}}~&{}{\sum\nolimits_{k=1}^{K}R_{k}}\label{40a}\\
\mbox{s.t.}~
&\text{(\ref{6c})}.&\label{40b}
\end{align}\label{40}%
\end{subequations}
The problem in (\ref{40}) can be solved by using the manifold optimization (MO) algorithm \cite{b14}. Finally, the BCD-based algorithm for solving problem (\ref{6}) is summarized in \textbf{Algorithm~3}. 
\begin{algorithm}[htbp]
\caption{BCD-based Algorithm for problem (\ref{6}).} 
\hspace*{0.02in}{\bf Initialization:} $b^{(0)}$, $\boldsymbol{w}^{(0)}$, $\boldsymbol{u}_{k}^{(0)}$, $\boldsymbol{\Theta}^{(0)}$.\\
\hspace*{0.02in}{\bf Repeat}\\
Update $b^{(j)}$ and $\boldsymbol{w}^{(j)}$ by using \textbf{Algorithm~1}.\\
Update $\boldsymbol{u}_{k}^{(j)}$ by using \textbf{Algorithm~2}.\\
Update $\boldsymbol{\Theta}^{(j)}$ by using the MO algorithm.\\
Update $j=j+1$.\\
\hspace*{0.02in}{\bf Until}
Convergence.\\
\hspace*{0.02in}{\bf Output:} 
$b^{*}$, $\boldsymbol{w}^{*}$, $\boldsymbol{u}_{k}^{*}$, $\boldsymbol{\Theta}^{*}$.\\
\end{algorithm}

\subsection{Complexity Analysis}
In this section, we compare the computational complexity of the proposed BCD-based algorithm with the following state-of-the-art algorithms.
\begin{itemize}
    \item Hybrid combining-based scheme with MO (SHC-MO): This is a codebook-based hybrid combination scheme, where the hybrid combiner is obtained by maximizing the system achievable rate. The RIS phase shifts are obtained by using the MO algorithm.
    \item Minimum mean-square error quantization bit allocations with MO (MMSQE-BA-MO): It is a variant of the algorithm in \cite{b19}, which optimizes the detection vectors of the users and the allocation of the quantization bits. The MO algorithm is used to obtain RIS phase shifts.
\end{itemize}
The total computational complexity of each iteration of the BCD-based algorithm is $\mathcal{O}(SM^{3.5}+K^{2.5}+KN_{t}+N_{r}N_{t})$. \textbf{Algorithm~3} has lower computational complexity than the SHC-MO algorithm, which amounts to $\mathcal{O}(M^{6}+KN_{t}+N_{r}N_{t})$ FPOS, and the MMSQE-BA-MO algorithm, which amounts to $\mathcal{O}(M^{6}+M^{2}N+KN_{t}^{2})$ FPOS. Therefore, our proposed BCD-based algorithm provides a better compromise between computational complexity and performance as confirmed by the numerical results illustrated in the next section.

\section{Numerical Results}\label{IV}
As shown in Fig.~\ref{fig:1}, we consider a single-cell system where $K=10$ users are uniformly distributed in the area $(2,30, 0)$~m and $(2,90,0)$~m. The simulation setup is $N_{t}=64$, $N_{r}=16$, $S=12$, $M=8$, $b^{\min}=1$, $b^{\max}=5$, $\sigma^{2}=-110$~dBm. The location of the AP is $(0,0,0)$~m and the location of the RIS is $(0, 40, 20)$~m.
The mmWave channels from the AP to the RIS, and from the RIS to the $k$th user are expressed as 
\begin{eqnarray}
{}{\boldsymbol{G}=\sqrt{1/\beta L_{1}}\sum\nolimits_{l=0}^{L_{1}-1}\alpha\boldsymbol{a}_{T}(N_{T},\theta_{l})\boldsymbol{a}_{R}^{T}(N_{r},\varphi_{l},\phi_{l}),}
\end{eqnarray}
\begin{eqnarray}
{}{\boldsymbol{h}_{k}=\sqrt{1/\hat{\beta}_{k}L_{k}}\sum\nolimits_{l=0}^{L_{k}-1}\hat{\alpha}_{k}\boldsymbol{a}_{R}(N_{r},\vartheta_{k}),}
\end{eqnarray}
where $\beta$ and $\hat{\beta}_{k}$ denote the large-scale fading coefficients. $L_{1}$ and $L_{k}$ denotes the number of paths between the AP and the RIS, and between the RIS and the $k$th user, respectively. $\beta$ and $\hat{\beta}_k$ are generated according to a complex Gaussian distribution \cite{b19,b20}
\begin{align}
\beta=\mathcal{CN}(0,10^{-0.1\kappa})\\
\hat{\beta}_{k}=\mathcal{CN}(0,10^{-0.1\kappa_{k}})
\end{align}
where $\kappa=72+29.2\log_{10}d+\zeta$ and $\kappa_{k}=72+29.2\log_{10}d+\zeta_{k}$. $d$ denotes the propagation distance, $\zeta\sim\mathcal{CN}(0,1)$ and $\zeta_{k}\sim\mathcal{CN}(0,1)$ account for the log-normal shadowing \cite{b1}, $\alpha$ and $\hat{\alpha}_{k}$ denote the small-scale fading coefficients whose distribution is $\mathcal{CN}(0,1)$ \cite{b10}. Also, $\boldsymbol{a}_{T}(\cdot)$ and $\boldsymbol{a}_{R}(\cdot)$ denote the array steering vectors at the AP and RIS, respectively.
In addition to the SCH-MO and MMSQE-BA-MO schemes, we also compare the following two schemes with \textbf{Algorithm~3}:
\begin{itemize}
\item Power allocation and hybrid combining with MO (PHC-MO): It is a variant of [18], which is implemented by a phase shifter, while considering the optimization of the user digital combiner.
\item NO-RIS: In this scheme, the RIS is not used. However, the  quantization bits and the beam selection are optimized by using \textbf{Algorithm~1} and \textbf{Algorithm~2}.
\end{itemize}

We investigate the convergence behavior of \textbf{Algorithm~3} in Fig.~2. We observe that the algorithm converges quickly, and, in general, only a few iterations are needed for ensuring the convergence. This shows that the proposed algorithm has low complexity. Fig.~3 exhibits the achievable sum rate (ASR) as a function of the number of reflecting elements of the RIS. We observe that the proposed algorithm achieves the best ASR. Although the resolution of the ADC for the NO-RIS scheme is higher than that of the RIS-aided schemes, we observe that the proposed BCD-based algorithm yields the same ASR as the NO-RIS scheme when the number of reflecting elements is $N_{r}=12$. This  finding validates the feasiblity of using RISs to mitigate the impact of low-resolution ADCs.
\begin{figure*}[htpb]
\centering
\subfigure{
\begin{minipage}[t]{0.25\linewidth}
\centering
\includegraphics[scale=0.27]{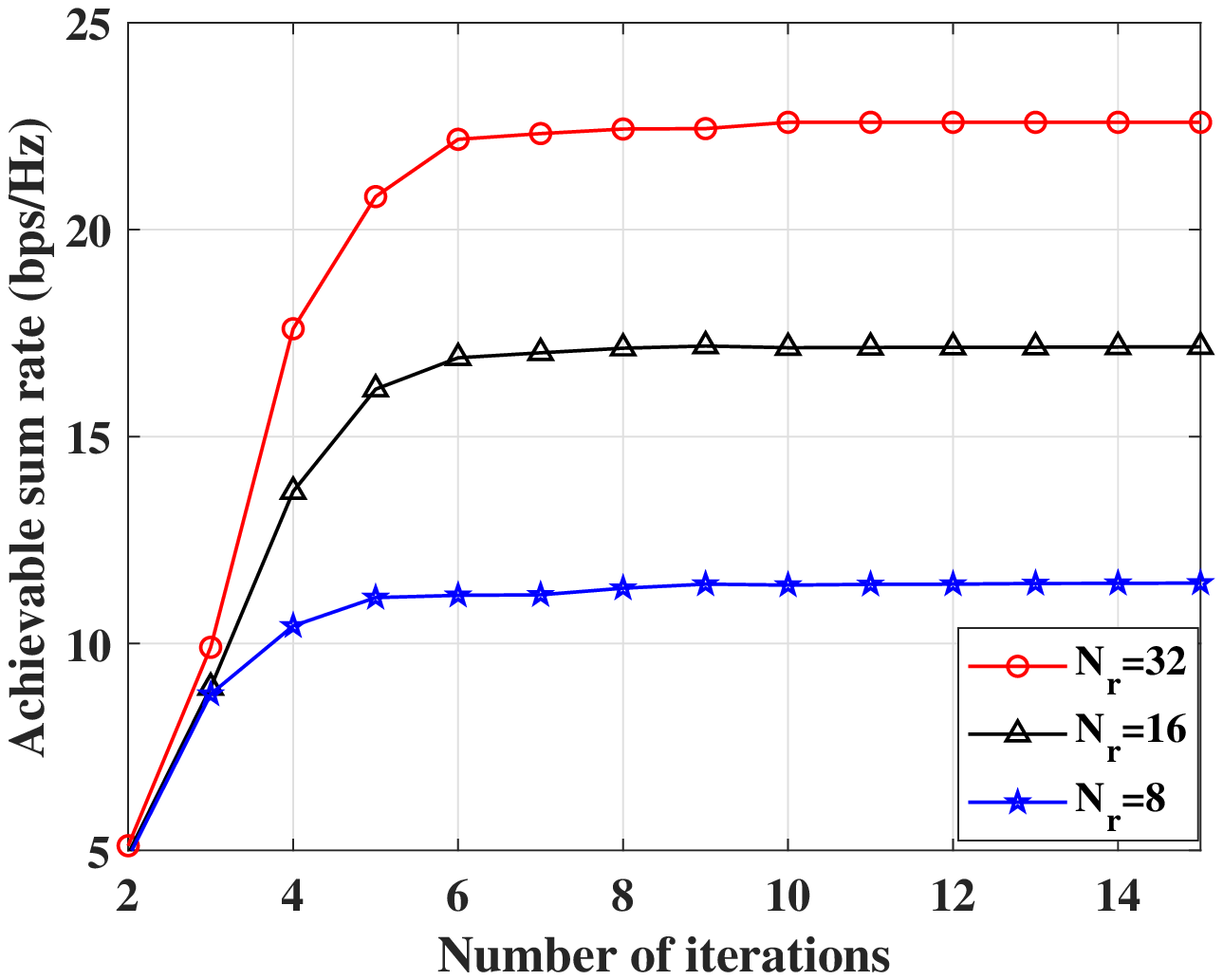}
\caption{Convergence}
\end{minipage}%
}%
\subfigure{
\begin{minipage}[t]{0.25\linewidth}
\centering
\includegraphics[scale=0.27]{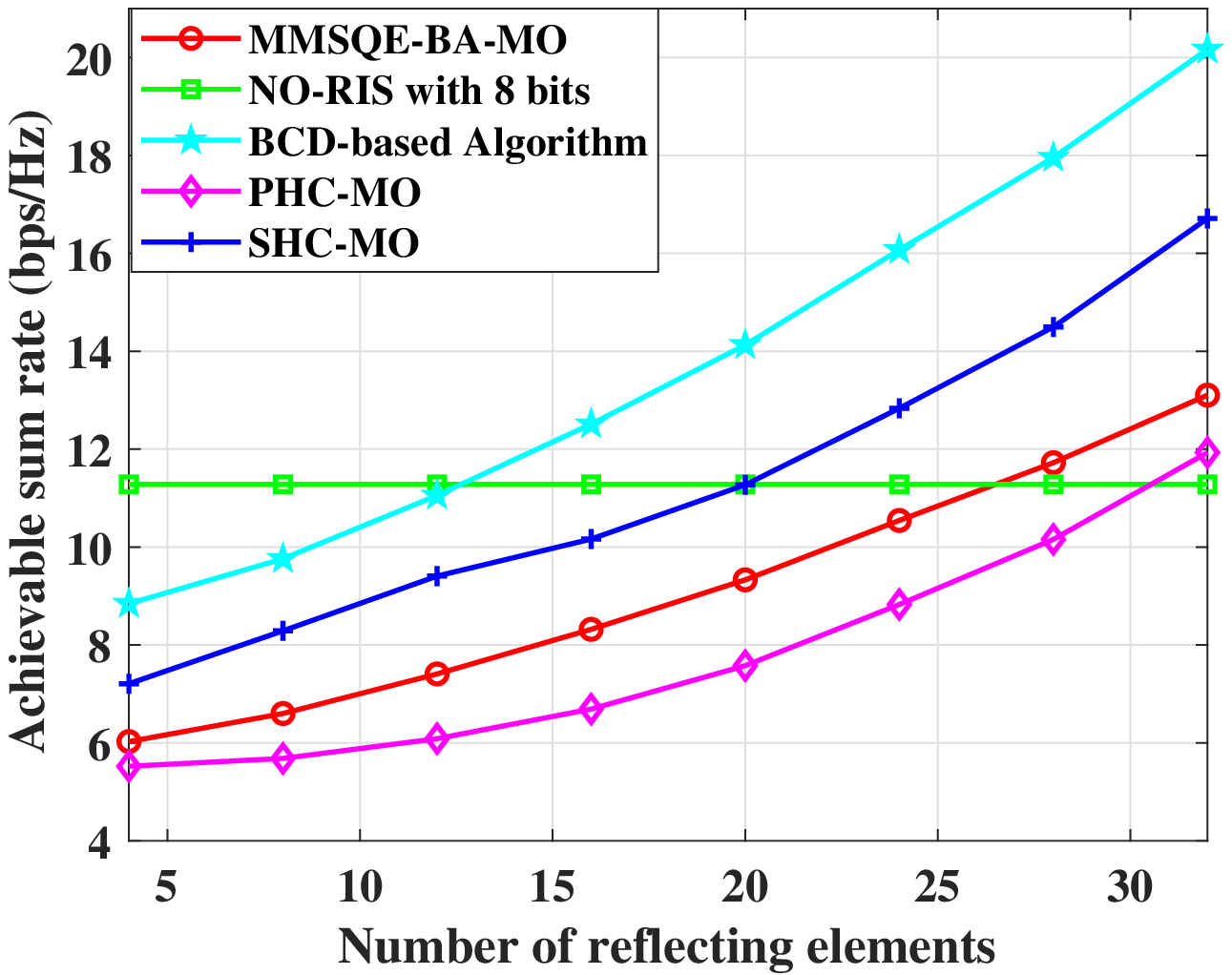}
\caption{ASR versus $N_{r}$.}
\end{minipage}%
}%
\subfigure{
\begin{minipage}[t]{0.25\linewidth}
\centering
\includegraphics[scale=0.27]{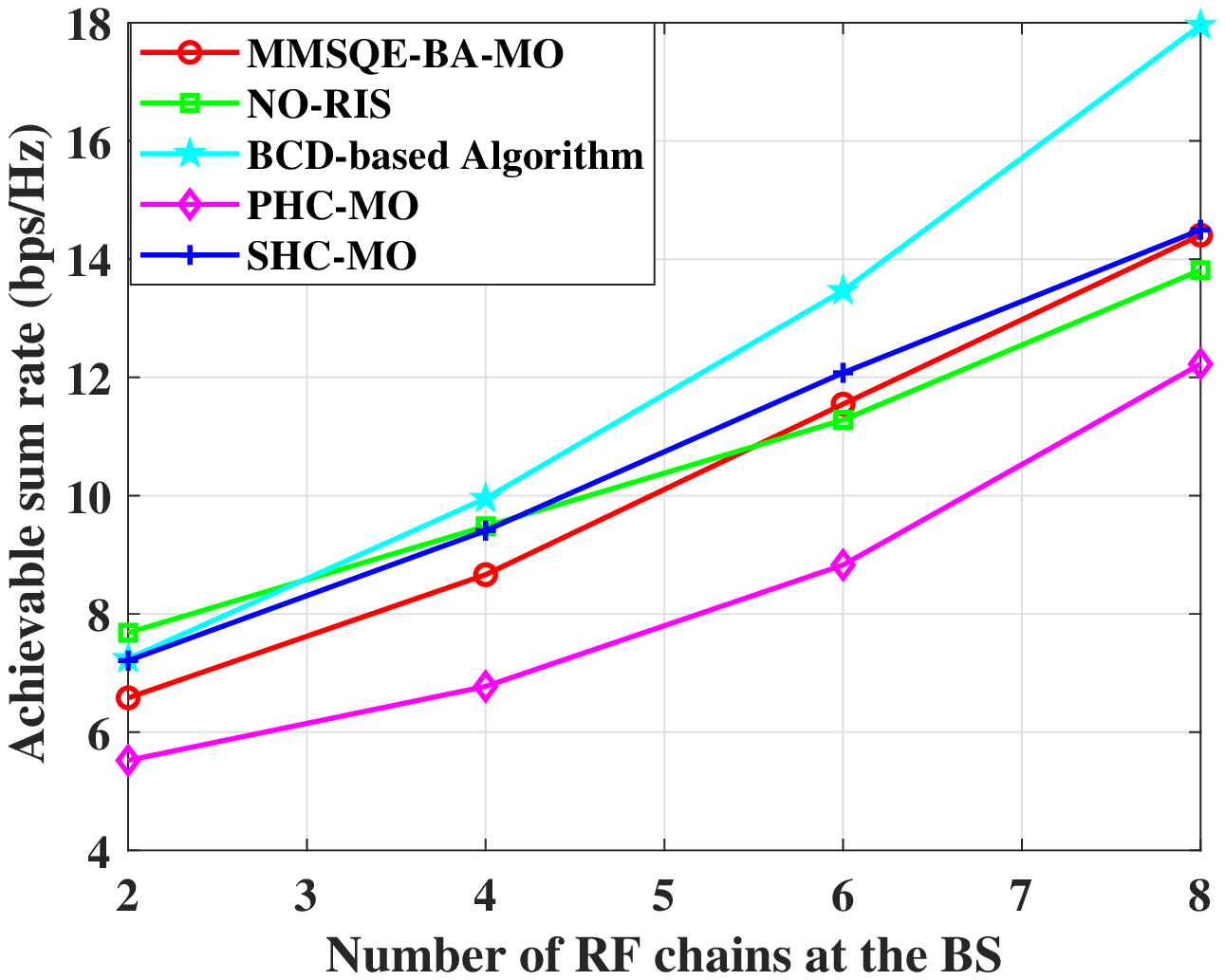}
\caption{ASR versus $M$.}
\end{minipage}
}%
\subfigure{
\begin{minipage}[t]{0.25\linewidth}
\centering
\includegraphics[scale=0.27]{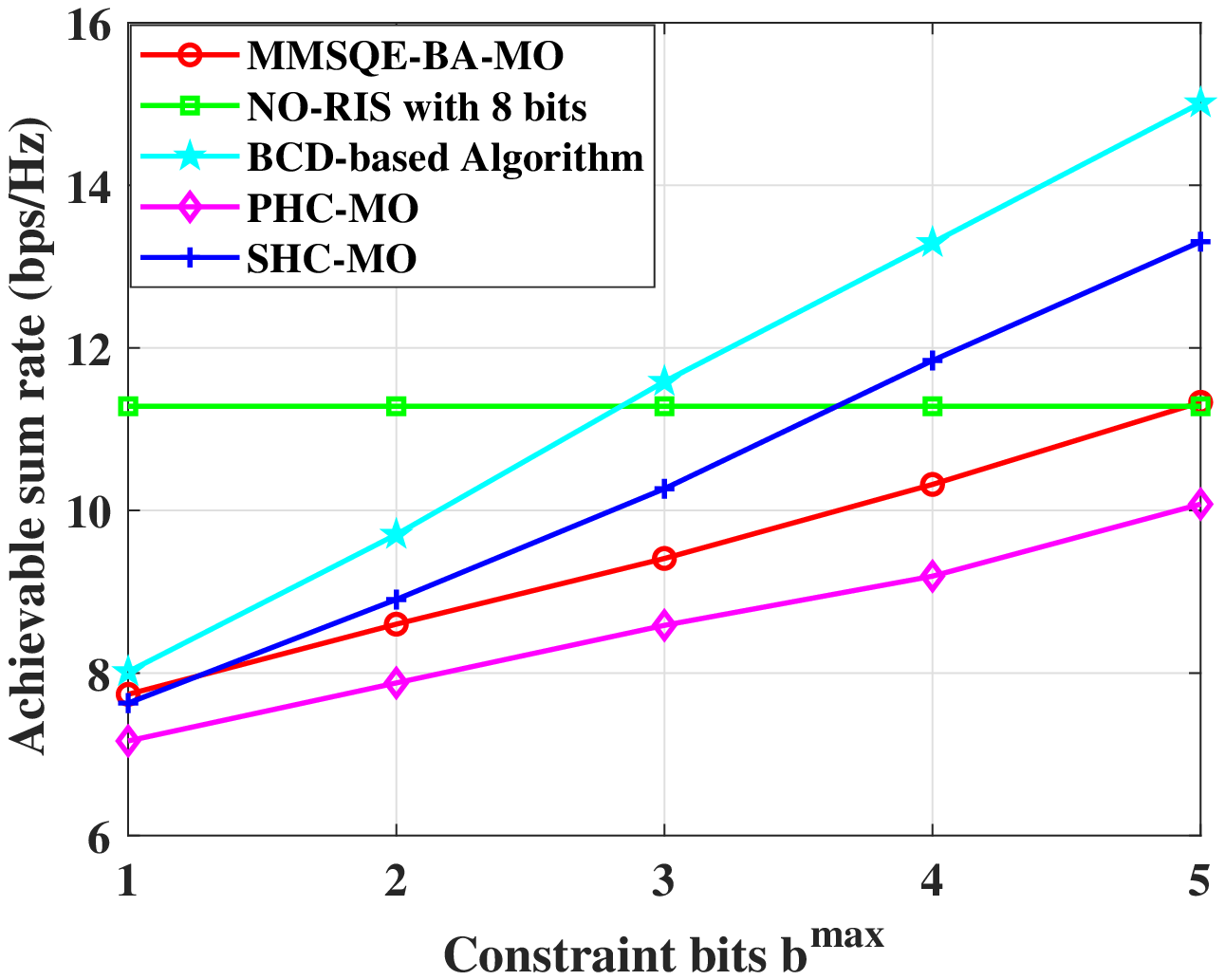}
\caption{ASR versus $b^{\max}$.}
\end{minipage}
}%
\centering
\end{figure*}

Fig.~4 shows the ASR as a function of the number of RF chains. Compared with all the other competing schemes, the proposed BCD-based algorithm yields the best ASR. From Fig.~4, we observe that the ASR increases with $M$. This is because a large number of RF chains can yield a high signal gain and can mitigate the interference. In addition, although the resolution of the ADC in the NO-RIS scheme is high, when the RF chains are $4$ and $6$, the proposed scheme and the conventional SHC-MO and MMSQE-BA-MO with RIS outperform the NO-RIS scheme. The result shows that there is no need to user a large number of RF chains when the system is aided by an RIS. 

Fig.~5 shows the ASR versus the maximum number of quantization bits for different schemes. Once again, we observe that the proposed BCD-based algorithm yields the best ASR. If $b^{\max}=3$, in particular, the proposed RIS-aided scheme outperforms the NO-RIS scheme with $8$ bits. This result highlights that the use of RISs can alleviate the impact of quantization noise.

\section{Conclusion}
The uplink achievable rate optimization problem for mmWave communications with hardware limitations at the AP was investigated, by jointly optimizing the RIS phase shifts, the beam selection matrix, decoding vector, and the quantization bits to maximize the sum rate. To deal with the resulting non-convex problem, a BCD-based algorithm is proposed. Simulation results showed that the proposed algorithm outperforms conventional algorithms in terms of ASR.


\ifCLASSOPTIONcaptionsoff
  \newpage
\fi


\bibliographystyle{IEEEtran}

\begin{thebibliography}{00}
\bibitem{b1} S. Rangan, T. Rappaport, and E. Erkip, ``Millimeter-wave cellular
wireless networks: Potentials and challenges,'' \emph{Proceedings of the
IEEE}, vol. 102, no. 3, pp. 366–-385, Mar. 2014.

\bibitem{b2} M. Di Renzo, ``Stochastic geometry modeling and analysis of multi-tier millimeter wave cellular networks,'' \emph{IEEE Trans. Wireless
Commun}., vol. 14, no. 9, pp. 5038–-5057, Sep. 2015.

\bibitem{b3}
M. Di Renzo, A. Zappone, M. Debbah, M.-S. Alouini, C. Yuen, J. de
Rosny, and S. Tretyakov, ``Smart radio environments empowered by
reconfigurable intelligent surfaces: How it works, state of research, and
road ahead,'' \emph{ arXiv preprint arXiv:2004.09352}.

\bibitem{b3_1}
M. Di Renzo et al., ``Reconfigurable intelligent surfaces vs. relaying:
Differences, similarities, and performance comparison,'' \emph{IEEE Open J. Commun. Soc}., vol. 1, pp. 798--807, 2020.

\bibitem{b6} S. Zhou, W. Xu, K. Wang, M. Di Renzo, and M.-S. Alouini, ``Spectral and energy efficiency of IRS-assisted MISO
communication with hardware impairments,'' \emph{IEEE Wireless Commun. Lett}., Early Access. 2020.

\bibitem{b9_2}
N.~Perović, L~Tran, and M.Di Renzo M, ``Achievable Rate Optimization for MIMO Systems with Reconfigurable Intelligent Surfaces,'' \emph{arXiv:2008.09563}.

\bibitem{b9_1}
 N. S. Perovic et al., ``Channel capacity optimization using reconfigurable
intelligent surfaces in indoor mmWave environments,'' in \emph{IEEE ICC(2020)}, pp. 1--7.

\bibitem{b10} C. Pradhan, A. Li, L. Song, B. Vucetic, and Y. Li, ``Hybrid precoding design for reconfigurable intelligent surface aided
mmwave communication systems,'' \emph{IEEE Wireless Commun. Lett}., vol. 9, no. 7, pp. 1041--1045, Jul. 2020.

\bibitem{b11} J. Choi et al., ``Resolution-adaptive hybrid MIMO architectures for
millimeter wave communications,'' \emph{IEEE Trans. Signal Process}., vol. 65,
no. 23, pp. 6201--6216, Dec. 2017

\bibitem{b13}
J. Song, P. Babu, and D. P. Palomar, ``Optimization methods for
designing sequences with low autocorrelation sidelobes,'' \emph{IEEE Trans. Signal Process}., vol. 63, no. 15, pp. 3998--4009, Aug. 2015.

\bibitem{b14}
P.-A. Absil, R. Mahony, and R. Sepulchre, \emph{Optimization algorithms on matrix manifolds}.    Princeton University Press, 2010

\bibitem{b15}
S. Boyd and L. Vandenberghe, \emph{Convex Optimization}. Cambridge univer-sity Press, 2004

\bibitem{b16}
P. Dong, H. Zhang, and G. Li, ``Spatially correlated massive MIMO relay systems with low-resolution ADCs,'' \emph{IEEE Trans. Veh Technol}., vol. 69, no. 6, pp. 6541--6553, Apr. 2020.

\bibitem{b17}
I. Yildirim, A. Uyrus, and E. Basar, ``Modeling and analysis of reconfigurable
intelligent surfaces for indoor and outdoor applications in future
wireless networks,” IEEE Trans. Commun. (to appear), Nov. 2020.

\bibitem{b18}
E. Basar, M. Di Renzo, J. De Rosny, M. Debbah, M.-S. Alouini, and
R. Zhang, ``Wireless communications through reconfigurable intelligent
surfaces,'' \emph{IEEE Access}, vol. 7, pp. 116753--116773, 2019.

\bibitem{b19}
C. Jinseok, B.L. Evans, and A.n Gatherer, ``Resolution-adaptive hybrid MIMO architectures for millimeter wave communications,'' \emph{IEEE Trans. Signal Process}., vol. 65, no. 23, pp. 6201--6216, Jun. 2017.

\bibitem{b20}
E. Basar and I. Yildirim, ``SimRIS channel simulator for reconfigurable
intelligent surface-empowered communication systems,” in Proc. IEEE
Latin-American Conf. Commun., Nov. 2020.
\end{thebibliography}

\end{document}